\documentclass[aps,prl,twocolumn,floatfix,superscriptaddress]{revtex4-1}

\usepackage{amsmath}
\usepackage{graphicx,bm}
\usepackage[colorlinks,citecolor=blue]{hyperref}
\begin{document}

\title{Fingerprints of heavy element nucleosynthesis in the late-time
  lightcurves of kilonovae} 

\author{Meng-Ru~Wu}
\email{mwu@gate.sinica.edu.tw}
\affiliation{Institute of Physics, Academia Sinica, 
  Taipei, 11529, Taiwan}
\affiliation{Institute of Astronomy and Astrophysics, Academia Sinica, 
  Taipei, 10617, Taiwan}

\author{J.~Barnes}
\email{jlb2331@columbia.edu}
\altaffiliation{NASA Einstein Fellow}
\affiliation{Department of Physics and Columbia Astrophysics
  Laboratory, Columbia University, Pupin Hall, New York, NY 10027,
  USA} 

\author{G.~Mart\'inez-Pinedo}
\email{g.martinez@gsi.de}
\affiliation{GSI Helmholtzzentrum f\"ur Schwerionenforschung,
  Planckstra{\ss}e~1, 64291 Darmstadt, Germany}
\affiliation{Institut f{\"u}r Kernphysik
  (Theoriezentrum), Technische Universit{\"a}t Darmstadt,
  Schlossgartenstra{\ss}e 2, 64289 Darmstadt, Germany}

\author{B.~D.~Metzger}
\email{bdm2129@columbia.edu}
\affiliation{Department of Physics and Columbia Astrophysics
  Laboratory, Columbia University, Pupin Hall, New York, NY 10027,
  USA} 

\begin{abstract}
  The kilonova emission observed following the binary neutron star
  merger event GW170817 provided the first direct evidence for the
  synthesis of heavy nuclei through the rapid neutron capture process
  ($r$-process).  The late-time transition in the spectral energy
  distribution to near-infrared wavelengths was interpreted as
  indicating the production of lanthanide nuclei, with atomic mass
  number $A \gtrsim 140$.  However, compelling evidence for the
  presence of even heavier third-peak ($A \approx 195$) $r$-process
  elements (e.g., gold, platinum) or translead nuclei remains elusive.
  At early times ($\sim$ days) most of the $r$-process heating arises
  from a large statistical ensemble of $\beta$-decays, which
  thermalize efficiently while the ejecta is still dense, generating a
  heating rate that is reasonably approximated by a single power-law.
  However, at later times of weeks to months, the decay energy input
  can also possibly be dominated by a discrete number of
  $\alpha$-decays, $^{223}$Ra (half-life $t_{1/2} = 11.43$~d),
  $^{225}$Ac ($t_{1/2} = 10.0$~d, following the $\beta$-decay of
  $^{225}$Ra with $t_{1/2} =14.9$~d), and the fissioning isotope
  $^{254}$Cf ($t_{1/2} = 60.5$~d), which liberate more energy per
  decay and thermalize with greater efficiency than beta-decay
  products.  Late-time nebular observations of kilonovae which
  constrain the radioactive power provide the potential to identify
  signatures of these individual isotopes, thus confirming the
  production of heavy nuclei.  In order to constrain the bolometric
  light to the required accuracy, multi-epoch and wide-band
  observations are required with sensitive instruments like the James
  Webb Space Telescope.  In addition, by comparing the nuclear heating
  rate obtained with an abundance distribution that follows the Solar
  r abundance pattern, to the bolometric lightcurve of AT~2017gfo, we
  find that the yet-uncertain r abundance of $^{72}$Ge plays a decisive
  role in powering the lightcurve, if one assumes that GW170817 has
  produced a full range of the Solar r abundances down to mass number
  $A\sim 70$.
\end{abstract}

\date{\today}
\maketitle


{\bf Introduction--}\label{sec-intro}
The gravitational wave emission detected from the binary neutron star
merger (NSM) GW170817 by Advanced LIGO \cite{Abbott.Others:2017}
triggered a world-wide search for electromagnetic counterparts
\cite{LIGO+17CAPSTONE}.  Within eleven hours of the coalescence, a
fading blue thermal source, AT~2017gfo, was discovered from
the galaxy NGC 4993~\cite{Coulter.Others:2017,Soares-Santos.Others:2017}.  
The luminosity and evolution agreed with predictions for the light powered
by the radioactive decay of heavy nuclei synthesized via the rapid
neutron capture process ($r$-process) in neutron-rich merger
ejecta~\cite{Li.Paczynski:1998,Metzger.Martinez-Pinedo.ea:2010,Roberts.Kasen.ea:2011,Barnes.Kasen:2013}.
The presence of luminous visual wavelength (``blue'') emission at
early times was interpreted by most groups as arising from the fastest
outer layers of the ejecta, which contained exclusively light
$r$-process nuclei with a relatively low visual wavelength
opacity~\cite{Metzger.Fernandez:2014,Nicholl.Others:2017,Drout.Others:2017}
(see, however, Ref.~\citep{Waxman+17,Kawaguchi:2018ptg}).
The observed transition of the emission colors to the near-infrared
confirmed predictions for the inner ejecta layers containing
lanthanide elements, with atomic mass number
$A \gtrsim
140$~\cite{Kasen.Badnell.Barnes:2013,Barnes.Kasen:2013,Tanaka.Hotokezaka:2013}.
The amount of the merger ejecta was estimated to be 
$M_{\text{ej}} \approx
0.03-0.06$~M$_{\odot}$~\cite{Kasen.Metzger.ea:2017, Kasliwal2017,Cowperthwaite.Others:2017,Villar2017,Waxman+17,Kawaguchi:2018ptg}, 
with the bulk of which expanding at velocities of $v_{\text{ej}} \approx 0.1$~c.

Although evidence exists for the presence of some lanthanides
in the ejecta of GW170817, the detailed abundance pattern of the
nuclei synthesized, and how it compares to those in the Solar System
or metal-poor stars, remains less clear.  This uncertainty arises
partly because of incomplete atomic data for the relevant elements and
ionization states, as well as the modeling of radiative transfer.
Even with accurate modeling, most kilonova properties at early times
$\sim 1$--10~days, when the lightcurves are at their peaks, 
are insensitive to the presence of even heavier nuclei, such as the
third-peak ($A \approx 195$) $r$-process elements (e.g., gold,
platinum) and transuranic nuclei.  Lanthanides are only produced in
ejecta with low electron fraction, $Y_{e} \lesssim
0.25$~\cite{Metzger.Fernandez:2014,Lippuner.Roberts:2015}, while even
smaller $Y_{e}$ are needed to synthesize heavier isotopes.  Whether
the ejecta of GW170817 contained such low $Y_{e}$ matter is presently
unknown.
 
At times after $\sim 10$~days, 
the ejecta becomes transparent, entering a ``nebular''
phase in analogy with those of supernovae, which are observed starting months after explosion.  
Although the uncertainties associated with the
ejecta opacity become smaller as it dilutes,  
these are replaced by even larger uncertainties in 
calculating the nebular spectrum, due to the increasing 
importance of deviations from local thermodynamical equilibrium
(see Ref.~\cite{Jerkstrand17} for a review in the supernova context).
Nevertheless, 
if one could measure the \emph{bolometric} nebular
emission, it should faithfully
track the radioactive decay energy input.

Table~I in the Supplemental Material (SM) lists \emph{all} 25 $r$-process isotopes with half-lives of $10-100$
days that can contribute to late-time heating.
Given the small number of isotopes, one might hope to detect 
the decay signatures of individual isotopes
and their associated yields, in the way that the $^{56}$Ni to
$^{56}$Co chain is observed in normal supernovae.  
As we shall show, these signatures could provide 
useful diagnoses of the range of heavy nuclei that are 
produced or even 
the elusive definitive proof that the heaviest nuclei in the 
universe are synthesized in NSM.

{\bf Late-Time Kilonova Heating--}\label{sec-LTKN}
We first examine the late-time kilonova emission for a few ejecta models that 
contain distinct nuclear compositions, as listed in Table~\ref{tab:model}.
In each model, the total \mbox{$r$-process} heating rate $\dot Q$ in the 
ejecta of total mass $M_{\text{ej}}$ and average expansion velocity
$v_{\text{ej}}$ can be formulated as
\begin{equation}\label{eq:dotQ}
\dot Q(t)=\sum_i f_i(t) \dot q_i(t) M_{\text{ej}}.
\end{equation}
It roughly equals the bolometric luminosity, $L_{\text{bol}}$, of the kilonova
following its peak light, particularly at late-time
after the ejecta becomes optically-thin.  
In Eq.~\eqref{eq:dotQ},
$\dot q_i(t)$ is the radioactive decay energy release rate per unit
mass from a decay channel $i$, including $\beta^-$-decay, $\beta^+$-decay/electron capture, 
$\alpha$-decay and spontaneous fission.  The
thermalization efficiency $f_i(t)$ is 
defined by the ratio of the 
rate of the ejecta specific thermal energy increase 
to $\dot q_i(t)$
due to the thermalization of decay products.  
We assume that
the material contains a Gaussian $Y_e$ distribution, characterized by
a central value $Y_{e,c}$ and a width $\Delta Y_e$.  The corresponding
$\dot q_i(t)$ is calculated using an $r$-process nuclear reaction
network~\cite{Wu2016}.  We adopt $f_i(t)$ of $\beta^-$-decay products
based on detailed particle thermalization simulations~\cite{Barnes+16}
while model those of dominating individual nuclei based on the work of
Ref.~\cite{Kasen:2018drm}.  These represent an important improvement
when compared with recent works~\cite{Zhu:2018oay,Wanajo:2018wra}.
Detailed descriptions for the calculation of $\dot q_i(t)$ and
$f_i(t)$ are given in the SM.

\begin{table}[ht]
  \caption{Late-time kilonova models (see text for explanations).}
\label{tab:model}
  \begin{ruledtabular}
    \begin{tabular}{ccccccc}
 Model & $Y_{e,c}$ & $\Delta Y_e$ & $A_{\text{peak}}$ &
$M_{\text{ej}}(\text{M}_\odot)$ & $v_{\text{ej}}(c)$ & Nuc. Mass.\\ \hline 
   A   & 0.15 & 0.04 & 130 \& 195 & 0.040 & 0.1 & FRDM \\
   B   & 0.25 & 0.04 & 80 \& 130 & 0.040 & 0.1 & FRDM \\
   C   & 0.35 & 0.04 & 80 & 0.055 & 0.1 & FRDM \\
   D   & 0.45 & 0.04 & 60 & 0.030 & 0.1 & FRDM \\
   A1   & 0.15 & 0.04 & 130 \& 195 & 0.020 & 0.1 & DZ31 \\   
    \end{tabular}
  \end{ruledtabular}
\end{table}

\begin{figure}
  \centering \includegraphics[width=\linewidth]{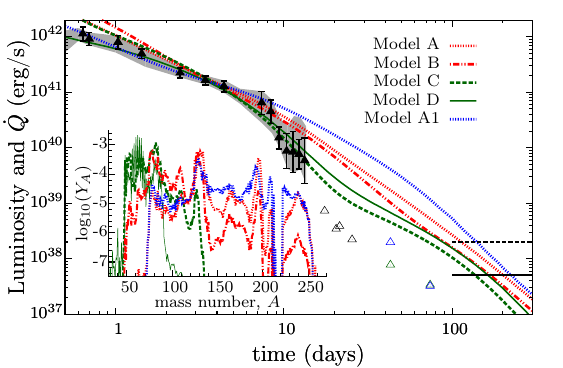}
  \caption{$L_{\text{bol}}$ of the kilonova associated with GW170817 from
    Ref.~\cite{Smartt.Others:2017} (filled black triangles), including
    uncertainties (grey band) derived from the range of values given
    in
    Ref.~\cite{Smartt.Others:2017,Cowperthwaite.Others:2017,Waxman+17}.
    Also shown are lower limits (empty triangles) on the late-time
    luminosity as inferred from the Ks band with
    VLT/HAWK-I~\cite{Tanvir:2017pws} (black) and the 4.5~$\mu$m
    detections by the \emph{Spitzer Space Telescope} from
    Ref.~\cite{Villar.Others:2018} (green) and
    Ref.~\cite{Kasliwal:2018fwk} (blue).  Colored lines show the
    ejecta heating rate $\dot Q(t)$ for different models listed in
    Table~\ref{tab:model}.  Their corresponding abundance
    distributions at $t=1$~d are shown in the inset.  The black solid
    (dashed) horizontal lines in the lower right corner represent the
    approximate observation limits of the NIR (MIR) instruments on the
    JWST for a merger at 100~Mpc.
  \label{fig:heating_FRDM}}
\end{figure}

For models A--D, we vary the ejecta $Y_e$ distribution such that the
produced peak and range of nuclei are largely distinct (see
Table~\ref{tab:model} and Fig.~\ref{fig:heating_FRDM}).  Both model A
and B with lower $Y_{e,c}=0.15$ and 0.25 produce a wide range of
nuclei across the two corresponding abundance peaks, $A_{\text{peak}}$.
On the other hand, model C and D with higher $Y_{e,c}=0.35$ and 0.45
only produce a smaller range of nuclei around its $A_{\text{peak}}=80$
and 60.

Fig.~\ref{fig:heating_FRDM} shows the inferred $L_{\text{bol}}$
of AT~2017gfo and the heating rate $\dot Q(t)$ derived
with models A--D. We vary the $M_{\text{ej}}$ 
to match the normalization of the luminosity at $\sim 3$--6 days.
Note that as we focus on the bulk of the ejecta, 
we ignore the early time data which most likely originated from a 
fast-moving component with different composition and lower mass.  
Fig.~\ref{fig:heating_FRDM} shows clearly that the $L_{\text{bol}}$ evolution in models that produce 
broad ranges of nuclei (A \& B) starts to diverge from those
with narrow ranges (C \& D) at $\sim 7$~days.
In particular, the latter cases show a clear dip at $\sim 25$~days.
This difference originates from the number of nuclei that can decay on timescales greater than $\sim$~days in each model. Both model A \& B contain $\sim 10$ nuclear species
that can decay at late times between 10--100 days, 
such that at any given time $t$ one can find a nucleus with a
commensurate $\beta$-decay lifetime $t_{1/2} \sim t$ contributing to the heating.  
This leads to a late-time power-law behavior of $\dot Q(t)$~\cite{Metzger.Martinez-Pinedo.ea:2010,Korobkin+12}.

However, for models C \& D which only produce nuclei around 
their $A_{\text{peak}}$, the absence of nuclei with
$\beta$-decay lifetimes in the range 10--50 days 
for $70\leq A\leq 100$ (see Table~I in the SM) results 
in the observed light curve dips at $\sim 25$~days.
Note that in both cases,
the resulting $\dot Q(t)$ are compatible with
the $L_{\text{bol}}(t)$ of AT~2017gfo and cannot be ruled out by such 
comparison alone (c.f., Ref.~\cite{Rosswog+17} which assumed single-$Y_e$ models).
A well-measured $L_{\text{bol}}(t)$ for future
events covering 10--50 days can be used to infer the range
of nuclei being produced in NSM\@.
Therefore, it can provide complementary information about the nuclear 
composition, in addition to the inferred mass fraction 
of lanthanides and actinides derived from comparison to radiation transport 
models, due to their high opacities that results in the reddening of the 
spectra~\cite{Barnes.Kasen:2013,Tanaka.Hotokezaka:2013}.

Models A--D use the same set of nuclear reactions. 
Previous studies show that the choice of theoretical nuclear physics
inputs can affect significantly the kilonova
lightcurves~\cite{Barnes+16,Rosswog:2016dhy} for low $Y_e$ ejecta,
as the $r$-process involves extremely neutron-rich nuclei, whose key properties (masses,
$\beta$-decay half-lives,\ldots) are not yet experimentally measured.
Particularly important are the produced amount of translead nuclei
that can undergo $\alpha$-decays or spontaneous fission at $\gtrsim$~days.
As they release a relatively large amount of energy per decay and their
decay products thermalize more efficiently than those of $\beta$-decays,
they can dominate the heating even in trace amounts.
Here, we illustrate the nuclear physics impact 
using two sets of neutron-capture rates and their reverse photo-dissociation
rates~\citep{Mendoza-Temis.Wu.ea:2015} which employ, respectively,
nuclear masses from the Finite-Range-Droplet-Model
(FRDM)~\cite{Moller:1993ed} for models A--D and the Duflo-Zuker parameterization with
31 parameters (DZ31)~\cite{Duflo:1995ep} for model A1.

\begin{figure}
  \centering
  \includegraphics[width=\linewidth]{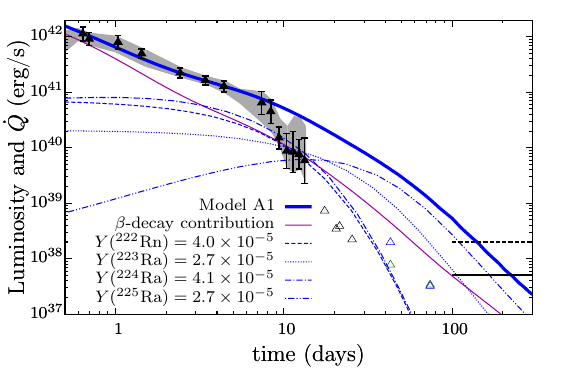}
  \caption{Lightcurve for the model A1 (thick
    solid blue line) showing the dominating contributions to the total
    radioactive heating: beta decays (thin maroon line) and individual
    $\alpha$-decays (thin blue 
    lines).\label{fig:heating-DZ}}
\end{figure}

Model A1 produces translead nuclei with $220\lesssim A\lesssim 230$ 
at the level of a few times $10^{-5}$, a factor of $\sim 4$--$10$ more than 
those by model A (see Fig.~\ref{fig:heating_FRDM}).
  Among those, four nuclei
  have $\alpha$-decay half-lives between 1 and 100 days:
  $^{222}$Rn($t_{1/2}=3.8$~days), $^{223}$Ra($t_{1/2}=11.4$~days),
  $^{224}$Ra($t_{1/2}=3.6$~days), and $^{225}$Ac($t_{1/2}=10$~days,
  following the $\beta$-decay of $^{225}$Ra with $t_{1/2}=14.9$~days).
Their decay chains
  release a large amount of nuclear energy $\sim 30$~MeV (see
  Table~I in SM), most of which goes
  into the kinetic energy of $\alpha$ particles, that thermalize
  more efficiently than $\beta$-decay products.  These
  $\alpha$-decays can therefore compete with the $\beta$-decays of
  many other nuclei at early time ($t\sim 2$--$6$~days) and 
  dominate the heating rate at late times, despite the
  abundances.  
We find that the enhanced heating 
from $\alpha$-decays reduces the required 
$M_{\text{ej}}$ to account for the AT~2017gfo luminosity around
  3-6 days by roughly a factor of 2 (see Table~\ref{tab:model}).  
  More importantly, it generates a
  broad ``bump''-like feature at $t \approx$ 6--200 days that is
  otherwise absent without actinide production.  This feature is
  mostly driven by the $A=225$ decay chain due to its effective long
  $t_{1/2}$ (see Fig.~\ref{fig:heating-DZ}).  
  As no other radioactive nuclei can release similar
  energy on this timescale, such a feature in future kilonova
  observations would uniquely point to the production of heavy nuclei
  up to the actinides in that mass range to the abundance level of a
  few times $10^{-5}$. 
  We also note that the steepening of the AT~2017gfo $L_{\text{bol}}$ at
    $t \sim 10$~d, places an upper limit of $\lesssim 10^{-5}$ for the
    total abundance of translead nuclei with $A=222$--225. 
This constraint may also used to derive upper limits 
on the U and Th production in GW170817 and future NSM
(see SM).

\begin{figure}
  \centering
  \includegraphics[width=\linewidth]{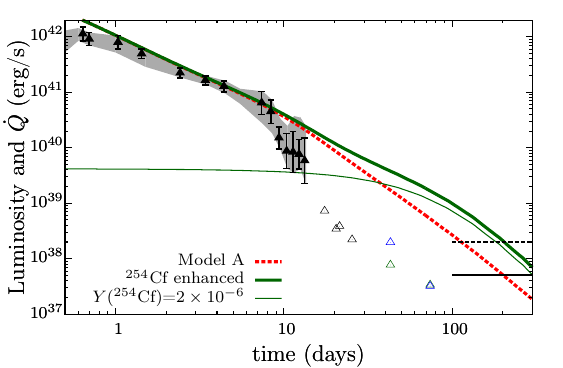}
  \caption{The decay of $^{254}$Cf could produce a late-time plateau
    in the lightcurve.  Thie figure is the same as
    Fig.~\ref{fig:heating_FRDM}, but showing both the model A and a case with the $^{254}$Cf abundance
    artificially enhanced.}
\label{fig:heating-Cf}
\end{figure}

Beyond the energy deposition from $\alpha$-decays, the
potential importance of spontaneous fission heating was pointed out in
Ref.~\cite{Wanajo+14} (also see Ref.~\cite{Zhu:2018oay} for a very
recent work discussing the impact of $^{254}$Cf fission on the
lightcurve).  Similar to the $\alpha$-decay nuclei, whether $^{254}$Cf
(or even heavier nuclei) can dominantly contribute to kilonova heating
is subject to nuclear physics uncertainties. The production of
$\alpha$-decay nuclei is sensitive to the evolution of the $N=162$
subshell closure for $Z\sim 80$ while the amount of $^{254}$Cf (and
neighboring nuclei) remaining at days is sensitive to the prediction
of fission barriers that affect various fission rates of the
progenitor nuclei~\cite{Vassh:2018wcf}.  Within our adopted nuclear
input, we do not find a significant contribution of $^{254}$Cf to the
heating rate when averaged over a wide range of $Y_e$ (see
Fig.~\ref{fig:heating_FRDM} for the low abundance of $A\gtrsim 250$).
Instead, we explore such an effect by artificially including a
fraction $Y(^{254}$Cf$) = 2\times 10^{-6}$ on top of the model A.
Fig.~\ref{fig:heating-Cf} shows that
even such a tiny quantity of $^{254}$Cf ($t_{1/2}=60.5$~days) produces
a lightcurve ``bump'' between 50--300 days.  We find that this feature
can be distinguished from that due to the late-time radioactive decay
of $^{56}$Co($t_{1/2}=77.24$~days), due to the very inefficient
thermalization of the $^{56}$Co decay products dominated by
$\gamma$-rays~\footnote{It is interesting to note that $^{254}$Cf was
  proposed to power the lightcurves of type Ia
  supernovae~\cite{Burbidge+56}}.
Note that a future identification of a 
``bump'' feature that does not match the timescale
by $\alpha$-decay or $^{254}$Cf 
fission discussed above may suggest the production of yet-unknown 
long-lived superheavy nuclei.

{\bf Heating from Solar r-abundances--}\label{sec-solarr}
One can ask whether the GW170817 kilonova
is consistent with that expected for ejecta containing $r$-process
nuclei with the Solar abundance pattern.  From detailed multi-band
lightcurve and spectral analyses, the inferred Lanthanide mass fraction, 
$X_{\text{lan}}$, is $\sim 10^{-3}$--$10^{-2}$~\cite{Kasen.Metzger.ea:2017,Tanaka:2017qxj,Waxman+17}.  
Assuming that the GW170817 yield follows the
Solar proportions, such low $X_{\text{lan}}$ requires the
production of 
all \emph{r}-process nuclei 
with additional contributions of trans-iron nuclei

We approach this question from the viewpoint of comparing the
luminosity of AT~2017gfo to the radioactive heating rate $\dot Q(t)$,
calculated under the assumption that the only heating contribution is
from $\beta$-decays and that the relative abundances of the unstable
nuclei follow exactly the Solar $r$-abundances ratios between
some minimum mass number $A_{\text{min}}$ and
$A_{\text{max}} = 205$~\cite{Tanaka+14}.  We employ two sets of the Solar
$r$-abundances from Ref.~\cite{Sneden.Cowan.Gallino:2008} (S1) and
Ref.~\cite{Goriely1999} (S2).

\begin{figure}
  \centering
  \includegraphics[width=\linewidth]{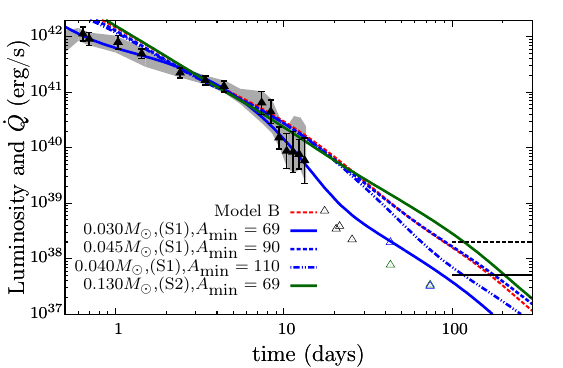}
  \caption{The radioactive decay heating rate powered by the
    Solar-$r$ abundance distribution for nuclei between
    $A_{\text{min}}$ and $205$. We use two different abundance sets
    from \cite{Sneden.Cowan.Gallino:2008,Goriely1999}.  See text for
    discussions.}
\label{fig:heating_sol}
\end{figure}

Fig.~\ref{fig:heating_sol} shows that with $A_{\text{min}}=90$ or $110$,
the resulting $\dot Q$ roughly matches $L_{\text{bol}}$ of AT~2017gfo for
$M_{\text{ej}}\simeq 0.04$~M$_\odot$.  In fact, they closely resemble the
model B prediction and both S1 and S2 give
consistent results.  However, such abundance patterns would have
$X_{\text{lan}}\gtrsim 0.1$, which is inconsistent with spectral modeling
of AT~2017gfo.

If we instead consider that GW170817 produced the Solar $r$-process
pattern down to $A_{\text{min}}=69$ (in order to
reduce $X_{\text{lan}}$ to values consistent with spectral modeling), for the S1 abundances the resulting $\dot Q$ can also be
consistent with the $L_{\text{bol}}$ of AT~2017gfo.  This model,
however, diverges from the $A_{\text{min}}=90$ or $110$ light curves beyond
10~d, a difference testable in future events.  
On the other hand, adopting the S2 abundances requires an uncomfortably 
large $M_{\text{ej}}\gtrsim 0.13$~M$_\odot$ to match the 
observed $L_{\text{bol}}$.  
This large difference arises because the abundance of $^{72}$Ge in S1 is similar 
to its neighboring nuclei, $^{70}$Zn and $^{74}$Ge, while for S2 the $^{72}$Ge abundance is zero.  
The only nucleus between $A=69-90$ that contributes significantly to 
the heating is the decay sequence, $^{72}$Zn ($t_{1/2}=1.94$~days) to $^{72}$Ga ($t_{1/2}=0.59$~days) 
to $^{72}$Ge, that releases a net energy $\sim 3.5$~MeV per decay.  The $\beta$-decay contribution of $^{72}$Zn in
S1 thus gives rise to the bump feature at 2--5 days that is lacking for the S2 set.
By artificially varying the $A=72$ mass fraction, we find that at least 
$\gtrsim 20$\% of its S1 abundance is needed match the GW170817 light curve for $M_{\text{ej}}\lesssim 0.05$~M$_\odot$
(see SM for details).

Taken together, we conclude that GW170817 may have produced a
solar-like $r$-process yield down to $A\sim 70$, if the solar
$r$-process contribution to the $^{72}$Ge abundance is 
larger than $\sim 20$\% of the value given by S1.
However, if 
the Solar $r$ abundance of $^{72}$Ge abundance turns out to be much smaller than 
that of $^{70}$Zn and $^{74}$Ge, then either a substantial
additional heating from $A<69$ isotopes (e.g., $^{66}$Ni,
see~\cite{Wanajo:2018wra}) would be required to make GW170817
consistent with the Solar abundances, or one would require enhanced
lighter nuclei yields in $A\sim 90-130$ relative to the
heavier nuclei beyond the second peak, when compared to the Solar
$r$-abundances, to give $X_{\text{lan}}\lesssim 0.01$.  
We note, however, that the correlation of 
the abundances of Ge and Fe in metal-poor stars and the non-correlation of Ge
and Eu \cite{Cowan:2005eh} hints that NSM are unlikely to produce
the entire solar $r$ abundances down to $A\approx 70$.

{\bf Discussion--}\label{sec-dis}
Our results demonstrate how late-time bolometric kilonova
  lightcurves can provide an important diagnostic of the nuclear
  composition of the NSM ejecta.  Recently,
Ref.~\cite{Villar.Others:2018,Kasliwal:2018fwk} reported detections of GW170817 at 43
and 74 days post-merger in the wavelength band centered at 4.5~$\mu$m
using the \emph{Spitzer Space Telescope}; the 3.6~$\mu$m band was also
observed, resulting in non-detections.  Interpreted as blackbody
emission, the observed colors indicate that the ejecta had cooled by
these late times to temperatures $\lesssim 1200$~K.  Unfortunately,
the ejecta during the nebular phase radiate through discrete
spectral lines rather than as a blackbody, and so translating these
observations into a bolometric luminosity is challenging.  Making the
very conservative assumption of counting only the luminosity 
in the detected band, these lower limits (shown as 
open triangles in Fig.~\ref{fig:heating_FRDM}--4) are not constraining in most of the cases.
The only exception is the scenario with heating powered by 
solar $r$-abundances with $A_{\text{min}}=69$ with the abundance set S1, for which 
the late-time lightcurve is in tension
with the data at 43 days of Ref.~\cite{Kasliwal:2018fwk}.

Observations of future merger events by, e.g., the James Webb Space
Telescope (JWST) could be more promising~\cite{Villar.Others:2018}.
For a merger at 100 Mpc, the NIRcam instrument on JWST could detect
luminosities in the $\approx 0.6$--4~$\mu$m band down to 
$L_{\text{NIR}} \approx 5\times 10^{37}$~erg~s$^{-1}$ (for a S/N = 10
detection given a $10^{4}$ s integration), sufficient to distinguish
various models shown in, e.g., Fig.~\ref{fig:heating_FRDM} out to
timescales of months.  The Mid-Infrared Instrument (MIRI) sensitive in
the 5--14~$\mu$m band, could constrain the luminosity to
$L_{\text{MIR}} \approx 2\times 10^{38}$~erg~s$^{-1}$.  We emphasize that
\emph{well time-sampled} observations, which cover as wide an
optical/infrared frequency range as possible, will be necessary to
constrain the bolometric lightcurve evolution with sufficient
precision to distinguish the nuclear physics features discussed here.

A number of uncertainties could affect future nebular measurements,
which requires additional theoretical modeling.  The ejecta may not
radiate the radioactive heating it receives with complete efficiency.
Empirically, the lightcurves of Type Ia supernovae faithfully track
the radioactive decay input up to several years~\cite{Kerzendorf+17}.
However, at later times the situation is less clear; non-thermally
excited ions might absorb a large fraction of the radioactive energy,
but due to the low density the rate of recombination could be slow and
the energy released much later than injection (``freeze-out'';
\cite{Fransson&Kozma93,*Fransson&Jerkstrand15}).  Freeze-out sets in
on timescales of years in supernovae (see Fig.~7 of
Ref.~\cite{Kerzendorf+17}), which, if occurring at the same density in
a NSM, would translate into an even earlier timescale of weeks to
months due to their lower ejecta mass and faster expansion speeds.

\begin{acknowledgments}
  The authors acknowledge useful discussions with Andrei Andreyev,
  Ben~Gibson, Karlheinz~Langanke, Karl-Heinz Schmidt and
  Friedrich-Karl~Thielemann, as well as anonymous referees for their helpful comments.  M.-R.W. acknowledges support from the
  Ministry of Science and Technology, Taiwan under Grant No.
  107-2119-M-001-038.  J.B. is supported by the National Aeronautics
  and Space Administration (NASA) through the Einstein Fellowship
  Program, grant number PF7-180162.  G.M.-P. is partly supported by
  the Deutsche Forschungsgemeinschaft (DFG, German Research
  Foundation) - Projektnummer 279384907 -
  SFB~1245. B.D.M. acknowledges support from NASA through the
  Astrophysics Research Program, grant number NNX16AB30G. Computations
  were partly performed on the LOEWE-CSC computer managed by Center for
  Scientific Computing of the Goethe University Frankfurt. We thank
  the ExtreMe Matter Institute EMMI at GSI, Darmstadt, for support in
  the framework of the EMMI Rapid Reaction Task Force ``The physics of
  neutron star mergers at GSI/FAIR'' during which this work has been
  initiated.
\end{acknowledgments}


%

%
\newpage
\clearpage

\onecolumngrid
\begin{center}
 {\em \large Supplemental Material}
\end{center}
\vspace*{0.2cm}
\twocolumngrid

{\bf Nuclei relevant for late-time kilonova lightcurves--} We list in
Table~\ref{tab:nuclei} all nuclei whose half-lives $t_{1/2}$ are
between $10$ and 100~days with atomic mass number $A>60$. There are in
total 25 isotopes and 22 decay sequences.  In addition, we include
$^{56}$Ni, $^{66}$Ni, $^{72}$Zn, $^{222}$Rn, and $^{224}$Ra 
despite the fact that their $t_{1/2}$ are smaller than 10 days.
This is because they can produce particular lightcurve features even at 
the early times, due to their relatively large energy release per decay.
For example, $^{56}$Ni and $^{66}$Ni are the main nuclei that contribute
to heating for model D at $t\lesssim 10$~days. 
In model C, it is instead mainly powered by the decay of $^{66}$Ni 
and $^{72}$Zn.

Among the listed nuclei, the four decay chains starting from $^{222}$Rn,
$^{224}$Ra, $^{223}$Ra and $^{225}$Ra contain 3--4 $\alpha$-decays and
release $\sim 20$--$30$~MeV decay energy in each.  If these nuclei can
be produced in an amount at the abundance level of $\sim 10^{-5}$,
they can generate distinct features in the late-time lightcurve between
$3-200$~days as discussed in the paper.  Moreover, the spontaneous
fission of $^{254}$Cf releases a even larger decay energy of
$\sim 185$~MeV. Therefore, a late-time feature shown in the kilonova
lightcurve at $t\gtrsim 50$~days can point to the production level of
$\sim 10^{-6}$.

We also note here again that in the list, there are no nuclei with
$10\ \text{days} <t_{1/2}<50$~days for $A<100$. If the merger ejecta
contains mostly high $Y_e$ material such that nuclei with this mass
range are primarily produced, a dip feature in the lightcurve around
$25$~days is expected to be identified as discussed in the paper. 

\begin{table*}[ht]
  \caption{The decay property of $r$-process nuclei with half-lives
    $t_{1/2} = 10-100$ days plus selected decays discussed in the main paper (from~\cite{nudat2decay}).  Nuclei that are 
    blocked by long-lived ($t_{1/2}\gg 100$~days) preceding isotopes
    are excluded.  $Q$ is the total energy released per decay (chain).
    $E_{\alpha}$, $E_{e}$, $E_{\gamma}$ are the total kinetic energy
    per decay (chain) carried by the $\alpha$, $e^\pm$ and photons,
    respectively.  For the spontaneous fission of $^{254}$Cf, the
    kinetic energy $E_{\text{Kinetic}}$ carried by the fission
    fragments is taken from Ref.~\cite{Brandt.Thompson.ea:1963}. No
    data is available for the neutron and photon effective energies
    but they are expected to be much smaller. \label{tab:nuclei}}
  \begin{ruledtabular}
    \begin{tabular}{ccccccc}
Isotope & Decay channel & $t_{1/2}$ & $Q$ &  $E_\alpha$ &   $E_e$ & $E_\gamma$\\ 
        &               & (d)      & (MeV) & (MeV)  & (MeV) & (MeV) \\
  \hline
  $^{56}$Ni & EC  & 6.075(10) & 2.133 & - & - & 1.721 \\ 
  $^{56}$Co & EC,$\beta^+$  & 77.236(26) & 4.567 & - & 0.121 & 3.607 \\ \hline
  $^{66}$Ni & $\beta^-$ to $^{66}$Zn & 2.2750(125) & 2.893 & - & 1.1396 & 0.098 \\ \hline\hline
  $^{72}$Zn & $\beta^-$  & 1.937(4) & 0.443 & - & 0.080 & 0.152 \\ 
  $^{72}$Ga & $\beta^-$  & 0.587(4) & 3.998 & - & 0.468 & 2.767 \\ \hline
  $^{224}$Ra & $\alpha\beta^-$ to $^{208}$Pb & 3.6319(23) & 30.875 & 26.542 &  0.891 & 1.474 \\ \hline
  $^{222}$Rn & $\alpha\beta^-$ to $^{210}$Pb & 3.8215(2) & 23.826 & 19.177 &  0.949 & 1.715 \\ \hline

  $^{225}$Ra & $\beta^-$ & 14.9(2) & 0.356 & - & 0.097 & 0.012 \\ 
  $^{225}$Ac & $\alpha\beta^-$ to $^{209}$Bi &  10.0(1) & 30.196 & 27.469 & 0.632 & 0.046 \\ \hline
  $^{246}$Pu & $\beta^-$ to $^{246}$Cm & 10.84(2) & 2.778 & - & 0.504  & 1.123\\ \hline
  $^{147}$Nd & $\beta^-$ & 10.98(1) & 0.895 & - & 0.232 & 0.144 \\ \hline
  $^{223}$Ra & $\alpha\beta^-$  to $^{207}$Pb & 11.43(5) & 29.986 & 26.354 & 0.937  & 0.304 \\ \hline
  $^{140}$Ba & $\beta^-$  to $^{140}$Ce &  12.7527(23) & 4.807 & - & 0.809 & 2.490 \\ \hline
  $^{143}$Pr & $\beta^-$ & 13.57(2) &  0.934 & - & 0.215 & - \\ \hline   
  $^{156}$Eu & $\beta^-$ & 15.19(8) &  2.452 & - & 0.430 & 1.235 \\ \hline
  $^{191}$Os & $\beta^-$ & 15.4(1) & 0.314 & - & 0.125 & 0.074 \\ \hline
  $^{253}$Cf & $\beta^-$ & 17.81(8) & 0.291 & - &  0.074 & - \\ 
  $^{253}$Es & $\alpha$ & 20.47(3) & 6.739 & 6.587 &  - & - \\ \hline
  $^{234}$Th & $\beta^-$ to $^{234}$U & 24.10(3) & 2.468 & - & 0.860 & 0.016 \\ \hline
  $^{233}$Pa & $\beta^-$ & 26.975(13) & 0.570 & - & 0.065 & 0.218 \\ \hline
  $^{141}$Ce & $\beta^-$ & 32.511(13) & 0.583 & - & 0.145 & 0.077 \\ \hline
  $^{103}$Ru & $\beta^-$ & 39.247(3) & 0.765 & - & 0.0638 & 0.497 \\ \hline
  $^{255}$Es & $\alpha\beta^-$ to $^{251}$Cf & 39.8(12) & 7.529 & 6.968 & 0.175 & 0.021 \\ \hline
  $^{181}$Hf & $\beta^-$ & 42.39(6) & 1.035 & - & 0.198 & 0.532 \\ \hline
  $^{203}$Hg & $\beta^-$ & 46.594(12) & 0.492 & - & 0.095 & 0.238 \\ \hline
  $^{89}$Sr & $\beta^-$ & 50.563(25) & 1.499 & - & 0.587 & 0.0 \\ \hline
  $^{91}$Y  & $\beta^-$ & 58.51(6) & 1.544 & - & 0.603 & 0.0 \\ \hline      
  $^{95}$Zr  & $\beta^-$ & 64.032(6) & 1.126 & - & 0.117 & 0.733 \\ 
  $^{95}$Nb  & $\beta^-$ & 34.991(6) & 0.926 & - & 0.043 & 0.764 \\ \hline      
  $^{188}$W  & $\beta^-$ to $^{188}$Os & 69.78(5)  & 2.469 & - & 0.878 & 0.061 \\ \hline      
  $^{185}$W  & $\beta^-$ & 75.1(3) & 2.469 & - & 0.127 & - \\ \hline\hline  
  Isotope & Decay channel & $t_{1/2}$ & Q & $E_{\text{Kinetic}}$ &   $E_n$ &
 $E_\gamma$ \\ 
        &  & (d) & (MeV)  & (MeV) & (MeV)  & (MeV)   \\ \hline
  $^{254}$Cf & Fission & 60.5(2) & - & 185(2)  & - & -
    \end{tabular}
  \end{ruledtabular}
\end{table*}

{\bf Modeling of the $r$-process and the radioactive decay--}
To model the \mbox{$r$-process} heating rate $\dot Q$ in the expanding
ejecta of total mass $M_{\text{ej}}$ and average expansion velocity
$v_{\text{ej}}$, we calculate the radioactive
decay energy release rate per unit mass from a decay channel $i$
(including $\beta^-$ decay, $\beta^+$ decay/electron capture,
$\alpha$ decay and spontaneous fission),
$\dot q_i$, assuming that the material contains a $Y_e$ distribution,
\begin{equation}\label{eq:qdot}
\dot q_i(t)=\int \dot q_i(t,Y_e) G(Y_e|Y_{e,c}, \Delta Y_e^2) dY_e,
\end{equation}
where $G(Y_e|Y_{e,c}, \Delta Y_e^2)$ is the normalized Gaussian distribution 
characterized by a central value $Y_{e,c}$ and a width $\Delta Y_e$.

As the $Y_e$-dependent heating rate $\dot q_i(t,Y_e)$ at the time
scale of $\sim 1-100$~days are completely determined by the abundance
distribution of nuclei produced during the $r$-process
nucleosynthesis, the abundances are computed by following the
evolution of all nuclear species from high temperature of
$\sim 10$~GK, when the nuclear composition is given by the nuclear
statistical equilibrium, to several Gyr, using an established
$r$-process nuclear reaction network (see e.g.,
Ref.~\cite{Mendoza-Temis.Wu.ea:2015,Wu2016}).

The reaction network contains all relevant reactions,
including charged particle reactions, neutron captures
and their inverse reactions, as well as the 
$\beta^-$ decays, $\alpha$ decays, $\beta^+$ decays/electron captures,
and the spontaneous, $\beta$-decay induced, and neutron-capture induced fission reactions. 
For all the theoretical reaction rates of neutron captures and the inverse photo-dissociations, 
$\beta^-$ decays, $\alpha$ decays, and fissions, we use those documented in~\cite{Mendoza-Temis.Wu.ea:2015}.
For the experimentally-known decay rates, we adopt the most-updated ones compiled by \cite{Audi:2017asy}.
Other reactions rates are taken from the JINA Reaclib Database of the Version v2.3~\cite{reaclib}.

\begin{figure}
\centering \includegraphics[width=\linewidth]{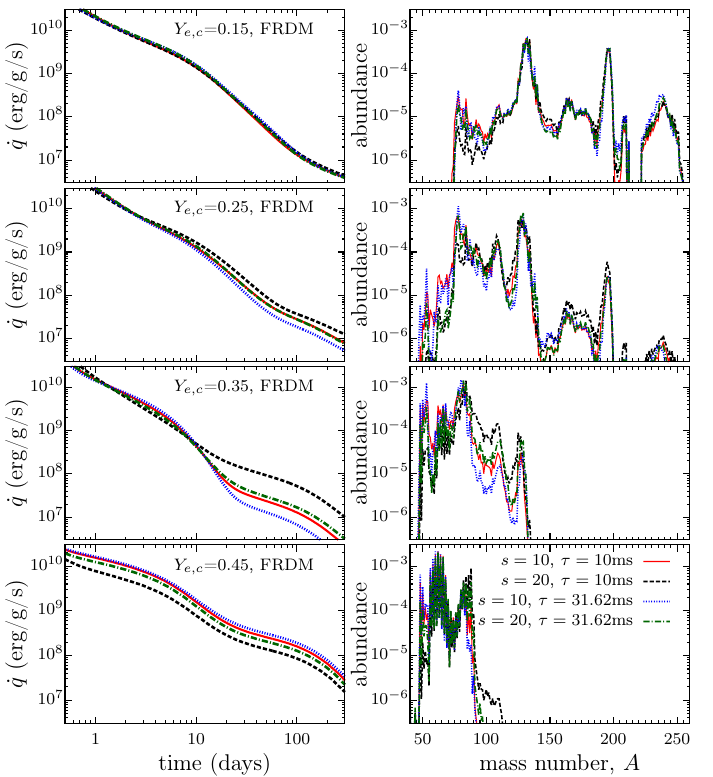}
\caption{The total specific rate of the radioactive energy release 
per unit mass $\dot q(t)\equiv\sum_i\dot q_i(t)$ (left panels) and
the $r$-process nucleosynthesis abundances (right panels)
for different $Y_e$ distributions
with various combinations of entropy $s$ and the 
early-time expansion timescale $\tau_{\text{dyn}}$.}
\label{fig:qd_comp}
\end{figure}

The expansion history of the ejecta used in the
$r$-process calculation is modeled by 
an analytically parametrized form
used in Ref.~\cite{Lippuner.Roberts:2015}, characterized by
the early-time expansion timescale $\tau_{\text{dyn}}$ and the 
entropy per nucleon $s$. For the results shown in the paper,
we use $\tau_{\text{dyn}}=10$~ms and $s=10$~$k_B$ per nucleon,
where $k_B$ is the Boltzmann constant.
The late-time shape of the heating rates does not 
sensitively depend on this particular choice of $\tau$ and $s$.
For example, Fig.~\ref{fig:qd_comp}
shows the total specific rate of the radioactive energy release 
per unit mass
$\dot q(t)\equiv\sum_i\dot q_i(t)$ for the combinations
of $(s[k_B],\tau_{\text{dyn}}[{\text{ms}}])=(10,10)$, $(10,31.62)$, $(20,10)$, and
$(20,31.62)$ for different $Y_e$ distributions
that correspond to the models in the paper.
It shows that despite the fact that the amount of late-time
radioactive energy release rate scales directly with the 
produced amount of nuclei which can decay over that timescale,
the shape remains in all cases.
This further illustrates the possibility
of using the late-time lightcurve shape to infer which specific nuclei
are present in the ejecta, as discussed in the paper.

{\bf Modeling of the particle thermalization--}
The fraction of energy from decay
products (photons, $e^\pm$, $\alpha$'s and fission fragments)
converted to heat (``thermalized'') in the ejecta depends on the bulk ejecta properties $M_{\text{ej}}$ and $v_{\text{ej}}$, and on the type of particle emitted, the emission spectrum of each particle type, and the rate at which the radioactivity produces energy.

We model the thermalization efficiencies of $\beta^-$ decay electrons and $\gamma$-rays by interpolating parametrized fits to the results of Ref.~\citep{Barnes+16}, which numerically calculated energy deposition assuming that the energy released by $\beta^-$ decay evolved in time as a power law $\dot{q}_{\beta}(t) \propto t^{-1.2}$. The numerical results were found to be well described by

\begin{equation}
f_{e^-}(t)=\frac{\ln(1+a t^b)}{a t^b},
\label{eq:ftb_elec}
\end{equation}
and 
\begin{equation}
f_{\gamma}(t)=1-\exp\left[-\left(\frac{t_{\gamma}}{t}\right)^d \right],
\label{eq:ftb_gam}
\end{equation}
where $t$ is the post-merger time in days.
The fit coefficients vary fairly smoothly with ejecta parameters, allowing them to be estimated for combinations of $M_{\text{ej}}$ and $v_{\text{ej}}$ not directly calculated by Ref.~\citep{Barnes+16}. 
Table~\ref{tab:betacoe} gives
$a$, $b$, $t_{\gamma}$, and $d$ for each $(M_{\text{ej}}, v_{\text{ej}})$ considered in this work.

\begin{figure}
\centering \includegraphics[width=\linewidth]{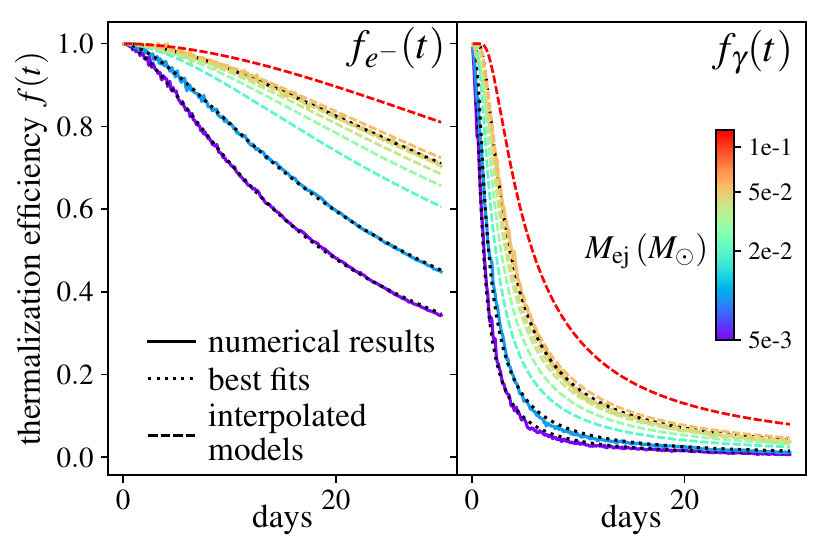}
\caption{Thermalization efficiencies $f(t)$ for $\beta$-decay
  electrons (left panel) and $\gamma$-rays (right panel). In both panels, solid, colored curves
  show numerical results from \cite{Barnes+16} with
  $v_{\text{ej}}=0.1c$, and $M_{\text{ej}}=0.005, \; 0.01, \; 0.05$, and
  $0.1$~M$_\odot$. The best-fit analytic expressions corresponding to
  Eqs.~\eqref{eq:ftb_elec} (left panel) and \eqref{eq:ftb_gam} (right
  panel) are over-plotted as dotted black lines to demonstrate the validity of the fitting function and the
  quality of the fit.  The
  thermalization efficiencies for the models in this work (see
  Table~\ref{eq:ftb_elec}) are estimated by interpolating fit
  coefficients as a function of ejecta parameters. The interpolated
  functions are plotted as dashed colored lines.  }
\label{fig:bgfits}
\end{figure}

Figure~\ref{fig:bgfits} presents numerical results for electron and
$\gamma$-ray thermalization from Ref.~\cite{Barnes+16} compared to the
best-fit analytic expressions calculated with Eqs.~\eqref{eq:ftb_elec}
and~\eqref{eq:ftb_gam} for select ejecta models similar to those
studied here. Also shown are the interpolated $f_{e^-}(t)$ and
$f_\gamma(t)$ for the models considered in this work. As shown in
Figure~\ref{fig:bgfits}, $f(t)$ evolves smoothly with bulk ejecta
properties, enabling confident interpolation of fit coefficients
between models.

\begin{table}[ht]
  \caption{Coefficients for the thermalization efficiency
  of $\beta$-decay electrons and $\gamma$-rays}
\label{tab:betacoe}
  \begin{ruledtabular}
    \begin{tabular}{cccccc}
$M_{\text{ej}}(\text{M}_\odot)$ & $v_{\text{ej}}(c)$ & $a(10^{-3})$ &  $b$ & $t_{\gamma}$
      (days) & d\\ \hline
  0.020       &    0.1 & 4.76 & 1.70 & 1.68 & 1.28 \\ 
  0.030       &    0.1 & 3.37 & 1.73 & 2.06 & 1.28 \\ 
  0.040       &    0.1 & 2.64 & 1.76 & 2.38 & 1.28 \\ 
  0.045       &    0.1 & 2.39 & 1.76 & 2.52 & 1.28 \\ 
  0.055       &    0.1 & 2.01 & 1.78 & 2.79 & 1.28 \\   
  0.130       &    0.1 & 0.97 & 1.84 & 4.28 & 1.28 \\ 
    \end{tabular}
  \end{ruledtabular}
\end{table}

The total instantaneous thermalization efficiency for
$\beta$-decays is then
\begin{equation}
f_\beta(t)=0.25 f_{e^-}(t) + 0.4 f_\gamma(t),
\label{eq:ft_beta}
\end{equation}
where the coefficients 0.25 and 0.4 approximate the partition of the
decay energy into $e^-$ and $\gamma$'s (with the remaining energy lost
to neutrinos, which escape the ejecta without thermalizing).  For the
results presented in this paper, we do not use the detailed branching
information provided in table~\ref{tab:nuclei}. However, this
information will be useful for future observations.

We find that the energy produced by $\alpha$-decays and fission is
dominated by a handful of decay chains, and therefore does not follow
a power-law.  Thermalization is sensitive to the form of $\dot{q}(t)$ \citep{Kasen:2018drm},
particularly at the late times we are probing in this paper.  Therefore, instead of adopting the results of
\citep{Barnes+16}, we have directly calculated $f_\alpha(t)$
and $f_{\text{fiss}}(t)$ for the \emph{individual} nuclei most important
for heating by these channels.
 
The procedure for these calculations is similar to that presented in
Ref.~\citep{Kasen:2018drm}. The energy-loss rates for $\alpha$
particles and fission were modeled as power laws
($\dot{E}_{\alpha, \text{fiss}} \propto E^\zeta$), where the power-law
index $\zeta$ and the coefficient of proportionality were chosen based
on the detailed energy-loss rates compiled by Ref.~\citep{Barnes+16}.
For $\alpha$-particles, we estimate
$\dot{E}_\alpha = \: 5 \times 10^{11}
\rho(t)$~MeV~s$^{-1} \: (\zeta=0)$, while for fission fragments we
find
$\dot{E}_{\text{fiss}} = 4.5 \times 10^{13} (E_{\rm
  fiss}/A)\times\rho(t)$~MeV~s$^{-1}$, where $A$ is the fragment's
mass number and $\rho$ is in units of g~cm$^{-3}$. In both cases, we
adopt a uniform density:
$\rho = 3M_{\text{ej}}/4\pi v_{\text{ej}}^3 t^3$.

These simplifications allow the analytic expression of a particle's
energy evolution with time, given its initial energy $E_0$, its birth
time $t_1$, and the current time $t$, The instantaneous deposition of
energy associated with any particular decay in a decay chain can be
solved numerically by calculating
\begin{align}
\dot{q}_{\text{dep}}(t) &= \int_{t_1}^t \dot{n}(t_1') \dot{E} (E(t, t_1')) {\text{d}} t_1',\label{eq:qdep1}
\intertext{with the associated thermalization efficiency given by}
f(t) &= \frac{\dot{q}_{\text{dep}}(t)}{\dot{q}_{\text{rad}}(t)} = \frac{\int_{t_1}^t \dot{n}(t_1') \dot{E} (E(t, t_1')) {\text{d}} t_1'}{E_0 \dot{n}(t)}.
\label{eq:qdep2}
\end{align}
In Eqs.~\eqref{eq:qdep1} and \eqref{eq:qdep2}, the lower limit $t_1$
is the birth time of the oldest particle not completely thermalized at
time $t$, and $\dot{n}(t)$ is the rate of particle emission.

In the case of $^{254}$Cf, $\dot{n}(t)$ is an exponential.
Because fission fragment thermalization is sensitive to the fragment's mass
and energy, we adopt a simplified model of the $^{254}$Cf fission fragment
distribution, in which every fission event
produces a heavy and a light fragment whose properties are the most
probable values measured by Ref.~\citep{Brandt.Thompson.ea:1963}. The
light (heavy) fragment has atomic number, mass number, and kinetic
energy 42, 109, and 102 MeV (56, 145, and 80 MeV), respectively. The
total $f(t)$ for $^{254}$Cf is an initial energy-weighted sum of $f(t)$
for each fragment.

For $\alpha$-decays, which generally occur as links in a longer
decay-chain and which do not exhibit exponential decay at all times,
we calculate number $N_i(t)$ of each nucleus in the chain, allowing
the determination of $\dot{n}_i(t) = -N_i(t)/\tau_i$, where
nucleus $i$ has a lifetime $\tau_i$. The thermalization efficiency for
the entire chain is simply a sum over its constituent $\alpha$-decays,
\begin{align}
f_{\rm tot,\alpha}(t) = \frac{\sum\limits_{i} f_i(t) E_{0,i}
  \dot{n}_i(t)}{\sum\limits_i E_{0,i} \dot{n}_i(t)}.  
\end{align}
Many of the decay chains that produce $\alpha$-particles also contain
nuclei that undergo $\beta^-$-decay. Because the
electrons emitted in these decays have energies that are comparable to
those of $\beta$-decay electrons from other \emph{r}-process nuclei
emitted at similar times, we consider these electrons as forming part
of the $\beta$-decay background, and absorb them into the general
calculation of $\beta$-decay thermalization (Eq.~\eqref{eq:ft_beta}).

\begin{figure}
\centering \includegraphics[width=\linewidth]{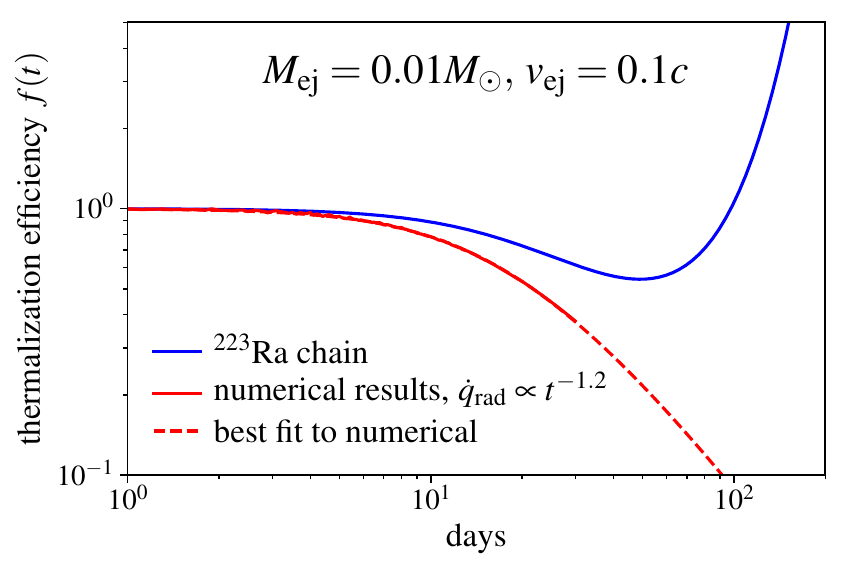}
\caption{The effect of $\dot{q}_{\text{rad}}$ on $f(t)$. The red curve
  shows the numerical results of \citep{Barnes+16} for
  $\alpha$-particle thermalization assuming power-law heating,
  $\dot{q}_{\text{rad}} \propto t^{-1.2}$.  An analytic fit to the
  numerical results, calculated using Eq.~\ref{eq:ftb_elec} (dashed
  red line), has been used to extend the curve past $t=30$ days. (The
  continual decrease of $f(t)$ for power-law heating is expected
  analytically, and is not imposed by our choice of fitting function.)
  For comparison, we plot in blue the thermalization efficiency
  calculated using Eq.~\ref{eq:qdep2} for the $\alpha$-particles
  produced by the decay chain originating with $^{223}$Ra, which shows
  qualitatively different behavior. The two curves begin to diverge
  around $t=5$ days, and the discrepancy increases with time.  }
\label{fig:aft_comp}
\end{figure}

Figure~\ref{fig:aft_comp} illustrates the impact of the heating rate
$\dot{q}_{\text{rad}}$ on the form of $f(t)$. We show, for an ejecta
model with
$(M_{\text{ej}}, v_{\text{ej}}) = (0.01\ \text{M}_{\odot}, 0.1\ c)$,
numerical results for $\alpha$ particles computed assuming a power-law
heating rate, compared to the semi-analytic $f(t)$ for a
representative $\alpha$-decay chain, $^{223}$Ra, calculated as
described above.

In general, the contribution of partially-thermalized particles emitted at earlier epochs causes $f(t)$ to decrease more slowly than $\dot{q}_{\text{rad}}(t)$. However, this effect is particularly strong for exponential decays, where the instantaneous energy deposition can actually exceed the instantaneous energy production, leading to $f(t) > 1$, as shown in Figure~\ref{fig:aft_comp}. 
The exponential (or
quasi-exponential) decay rates in the case of $^{223}$Ra result in a less
steep decline at intermediate times, and cause
$f(t) \rightarrow \infty$ as $t \rightarrow \infty$. 
While the position and depth of
the local minimum depend on ejecta parameters and the $Q$-values and
timescales of the decays in question, the asymptotic behavior is a
robust feature of single-isotope/single-decay chain heating.
Despite the asymptotic behavior of $f(t)$ in this regime,  $\dot q_{\text{rad}}(t)\times f(t)$ remains finite at all times and
asymptotes to a power-law as
$t\rightarrow\infty$~\cite{Kasen:2018drm}, and the time-integrated deposited energy is less than the total radiated energy.

In compositions neutron-poor enough to synthesize $^{56}$Ni, the decay
of the $^{56}$Ni daughter $^{56}$Co proceeds via $\beta^+$ decay with
a branching ratio of 19\%, creating a population of high-energy
positrons that carry $\sim3$\% of the total decay energy. Most of the
energy (79\%) is carried by $\gamma$-rays, with neutrinos accounting
for the remainder.  As with electrons, the primary channel for positron
energy loss is Bethe-Bloch
interactions~\cite{Chan.Lingenfelter.1993_positronsSNe}; however, as a
result of different relativistic corrections to the Bethe-Bloch
formula for electrons and positrons, the energy-loss rates for
positrons with energies near 1 MeV are slightly higher than those for
electrons.

We find that the energy-loss rate for positrons varies roughly as
$E^{-1/4}$ in the energy range of interest. We then calculate the
thermalization efficiency of positron energy, $f_{e^+}(t)$, as
described above for $\alpha$-particles and fission fragments. The
total thermalization efficiency of the
$^{56}\text{Ni} \rightarrow{} ^{56}\text{Co} \rightarrow{} ^{56}\text{Fe}$
chain is approximately
\begin{equation}
f_{\beta^+/\text{EC}}(t)=0.03 f_{e^+}(t) + 0.79 f_\gamma(t).
\end{equation}

{\bf Inferring U and Th abundances--} We discuss in this
  section the possibility of inferring the U and Th abundances
  utilizing the (non-)detection of the lightcurve signature of
  translead nuclei.  Taking GW170817 as an example, the produced
  abundance of $A=222$--225 nuclei can be estimated to be smaller than
  $\sim 10^{-5}$.  One can then take a model among those with varying
  composition and nuclear input, which predicts the largest
  ratio of nuclei that eventually decay to the U/Th abundances
  relative to the $A=222$--225 nuclei, to infer a ``model-dependent''
  upper limit for the amount of U and Th produced.
  Within the scenarios examined here, model A, based 
  on the FRDM nuclear masses, gives such a largest ratio with an
  total abundance of $\sim 10^{-5}$ (mass fraction
  $\sim 2.2\times 10^{-3}$) for $A=222$--225 nuclei, and the U/Th
  abundances (after decay) of $\sim 1.5\times 10^{-4}$ (mass fraction
  $\sim 3.5\times 10^{-2}$).  Therefore, the upper limit on the U/Th
  abundances can be set to be $\sim 3.5\times 10^{-2}$, which
  translates to a total amount of $\sim 1.75\times 10^{-3}$~M$_\odot$
  when taking $M_{\text{ej}}\sim 0.05$~M$_\odot$.

Similarly, a future (non-)observation of the feature of $^{254}$Cf
may likewise be used to determine the U/Th yield.

\begin{figure}
\centering \includegraphics[width=\linewidth]{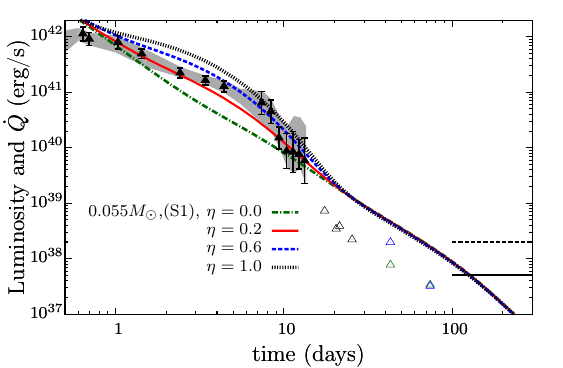}
\caption{Heating rate powered by the Solar-$r$
abundances distribution for nuclei between $69\leq A\leq 205$
from the abundance set S1~\cite{Sneden.Cowan.Gallino:2008}.
Due to the large uncertainty of $^{72}$Ge abundance in 
different abundance sets (see paper) and its dominating role
in heating, the amount of $A=72$ nuclei has been adjusted by
$Y(A=72)=\eta\times Y(A=72)|_{S1}$.}
\label{fig:ge72}
\end{figure}

{\bf Solar $r$-abundances of $^{72}$Ge and GW170817--} As discussed in
the paper, the amount of $^{72}$Ge in the Solar $r$-process
abundances, which is uncertain, plays a key role in connecting the
Solar $r$-process abundances with the GW170817 kilonova lightcurve.
Fig.~\ref{fig:ge72} further shows the comparison of the
$L_{\text{bol}}$ of AT~2017gfo with $\dot Q$ powered by the
$\beta$-decay of nuclei that follow the Solar $r$-process distribution
for $69\leq A\leq 205$, with the amount of
$Y(A=72)=\eta\times Y(A=72)|_{S1}$, for a fixed ejecta mass
$M_{\text{ej}}=0.55$~M$_\odot$.  It shows that even with such large
$M_{\text{ej}}$, it requires $\eta\gtrsim 0.2$ to power the observed
lightcurve, if we demand that GW170817 produces the entire range of
the $r$-process nuclei.

\end{document}